\documentclass[conference]{IEEEtran}

\usepackage{booktabs} 

\usepackage{enumitem}

\usepackage{comment} 

\usepackage{graphicx}
\usepackage{tabularx}
\usepackage{xcolor,comment,afterpage}
\usepackage{amsmath,amssymb,amsfonts}
\usepackage{algorithmic}
\usepackage{textcomp}
\usepackage{color,soul}
\usepackage{lipsum}  
\newcolumntype{L}{>{\raggedleft\arraybackslash}X}
\newcolumntype{R}{>{\raggedright\arraybackslash}X}
\usepackage{color}
\usepackage{graphicx}
\usepackage{tabularx}
\usepackage{hyperref}

\usepackage{multibib}
\newcites{P}{The systematic reviews in the tertiary study}

\begin{document}
\title{Experiences on Managing Technical Debt
with Code Smells and AntiPatterns}



\author{\IEEEauthorblockN{Jacinto Ramirez Lahti}
\IEEEauthorblockA{\textit{Solita Ltd, Finland} \\
jacinto.ramirez@solita.fi}
\and 
\IEEEauthorblockN{Antti-Pekka Tuovinen}
\IEEEauthorblockA{\textit{Department of Computer Science} \\
\textit{University of Helsinki, Finland}\\
antti-pekka.tuovinen@helsinki.fi}
\and
\IEEEauthorblockN{Tommi Mikkonen}
\IEEEauthorblockA{\textit{Department of Computer Science} \\
\textit{University of Helsinki, Finland}\\
tommi.mikkonen@helsinki.fi}
}

\maketitle              

\begin{abstract}
\emph{Technical debt} has become a common metaphor for the accumulation of software design and implementation choices that seek fast initial gains but that are under par and counterproductive in the long run. However, as a metaphor, technical debt does not offer actionable advice on how to get rid of it. To get to a practical level in solving problems, more focused mechanisms are needed. Commonly used approaches for this include identifying {\em code smells} as quick indications of possible problems in the codebase and detecting the presence of {\em AntiPatterns} that refer to overt, recurring problems in design. There are known remedies for both code smells and AntiPatterns.
In paper, our goal is to show how to effectively use common tools and the existing body of knowledge on code smells and AntiPatterns to detect technical debt and pay it back. We present two main results: (i) How a combination of static code analysis and manual inspection was used to detect code smells in a codebase leading to the discovery of AntiPatterns; and (ii) How AntiPatterns were used to identify, characterize, and fix problems in the software. The experiences stem from a private company and its long-lasting software product development effort. 
\end{abstract}

\begin{IEEEkeywords}
Technical debt, code smells, AntiPatterns, case study, software maintenance, code refactoring
\end{IEEEkeywords}

\section{Introduction}

The need for constant change has become a key driver in software development. Bug fixes, new features, and technology and architecture updates take place frequently, even multiple times per day \cite{feitelson2013development}. At times, changes have been carefully considered whereas at some other times, changes are made in a hurry to fix an urgent issue in the codebase.

Technical debt \cite{cunningham1992t} has become a common metaphor for the accumulation of software design and implementation choices that seek fast initial gains but that are under par and, consequently, counterproductive in the long run \cite{fraser2013technical}. A typical example of this would be risking the internal quality of the codebase in favor of delivering features faster. Over time, these deficiencies in code quality build up in a similar way as financial debt does if nothing is done to pay it back. Then, any new change will require ever more time and effort.

While technical debt is a useful concept for characterizing the status of a software system, it is often too generic to offer actionable advice, other than saying that the codebase should be improved or fixed. To get to a practical level in solving problems, more focused mechanisms are needed -- the metaphorical concept must be reified \cite{ernst2015measure}. Commonly used approaches for this include identifying code smells \cite{fowler2018refactoring} and detecting AntiPatterns \cite{brown1998refactoring}. The former refers to quickly identifiable indications of possible problems in the codebase; the latter refers to a manifest, recurring problem in design that can be fixed by following a prescribed sequence of changes.

In this experience report, our goal is to show how to effectively use common tools and the existing body of knowledge on code smells and AntiPatterns to detect technical debt and pay it back. We present two main results: (i) How a combination of static code analysis and manual inspection was used to detect code smells in a codebase leading to the discovery of AntiPatterns; and (ii) How AntiPatterns were used to identify, characterize, and fix problems in the software. The work is based on the lead authors'  Master's thesis \cite{jacinto-thesis}, where the full research setup is presented. Here, we focus on the practical part of the work, done for a private company.

The rest of this paper is structured as follows. In Section 2 we present the background of this work. In Section 3, we present an overview of the case company. In Section 4, we discuss the study we did in the company's context with respect to code smells and AntiPatterns. In Section 5, we list the lessons we learned in the process. Finally, in Section 6, we draw our conclusions.

\section{Background and Related Work}

Technical debt is a metaphor that reflects the implied cost of additional deferred rework caused by choosing an expedient, easy solution now instead of following a recognized best practice that is more laborious. While the root causes of occurrences of debt are usually the same -- solutions are designed and implemented without understanding the big picture or with limited resources and under time pressure -- its manifestations are many. For instance, legacy code and opportunistic designs \cite{hartmann2008hacking} whose behavior is hard to understand are practical symptoms of technical debt. However, missing tests and other quality assurance infrastructure issues that complicate verification can be regarded as technical debt, too.  

Due to the wide interpretation, it is difficult to be very precise when using the term. Hence, to understand and control technical debt, other concepts are needed. While the breakdown of technical debt has been proposed, including testing debt, design debt, defect debt, and documentation debt \cite{zazworka2013case,ernst2015measure}, even this results in concepts that require further operationalization (reification) for guiding concrete actions. 

In our case, we will use two additional concepts to manage technical debt. These are code smells, used in looking for technical debt, and AntiPatterns, used for characterizing debt and introducing a recipe for paying it back. These two concepts will be addressed next. 

\subsection{Code Smells}

A code smell has been defined as a surface indication that usually stems from a deeper problem in the system \cite{code-smell,fowler2018refactoring}. Not all smells end up being bad -- for example some long methods have a reason to exist -- but a long method is usually considered dubious and it should be investigated to corroborate that it is not. The above definition also indicates that usually the smelly part might not constitute the actual problem by itself but it might be a signal of deeper problems.

A key aspect of code smells is that they have one or more associated refactoring solutions that can help getting rid of the smell. For example, in the case of long functions, the {\em Extract Function} refactoring \cite[p. 106]{fowler2018refactoring} directs us to divide the logic of a long function into shorter functions.

One way to detect possible code smells automatically is to use static source code analysis tools \cite{emden2002java,almashfi2020js,zampetti2017open}. Static source code analysis tools analyze the source code without the need to compile and run it and there are different tools discerning different things in the code \cite{louridas2006static,spinellis2006bug,rios2019supporting}. Some tools, like Lint \cite{johnson1977lint}, are usually run continuously in the Integrated Development Environments (IDE) or in editors used to write code, which allows the tool to immediately warn the developer of suspicious code while it is being written. Tools can also provide visualized representations of the code and the relations between different files, classes and functions.

\subsection{AntiPatterns}

While code smells reflect symptoms that are quick to spot -- almost intuitively -- and there are several ways to rectify the underlying problems, AntiPatterns \cite{brown1998refactoring} are more formal in their nature. This formalization covers not only the problematic pieces of design but also their underlying causes and often also a generalized refactored solution for the problem. 
 
Since AntiPatterns not only describe problems but also potential solutions, they offer a way to harmonize the refactoring of a codebase. In other words, to fix a recurring problem, a repeatable solution is applied. This in turn keeps the codebase more uniform than following an approach where individual code smells are dealt with individually in various ways. In general, this helps in maintaining the codebase in the long term.

Some sources call AntiPatterns design smells. In fact, a recent systematic literature survey has recognized a number of inconsistent definitions for many kinds of `software smells' \cite{sharma2018survey}. However, in this report, we use the terms code smell and AntiPattern in their original meanings. We used \cite{fowler2018refactoring} as the catalog of code smells and  \cite{brown1998refactoring} as the catalog of AntiPatterns.

There is active research in the area of tools for detecting smells and AntiPatterns, but the results are somewhat mixed \cite{sharma2018survey} and not very well received by practitioners \cite{ernst2015measure}. For example, the lack of available tooling for identifying and measuring technical debt related to deep architectural structures of software is recognized as a problem in the empirical study \cite{yli-huumo2016tdm} of eight software development teams in a large organization that provides multiple software solutions. 

According to the study \cite{yli-huumo2016tdm}, the identification of technical debt relies mostly on developers discovering problematic code structures during development and on architects doing deeper analysis with the help of static analysis tools to mark suspect code. The study did not go into details about how analysis work and corrective actions (repayment) were planned and carried out but emphasized the manual work needed. Our study shows that using AntiPatterns to (i) characterize the deeper design problems and the solutions proposed by the AntiPatterns as guidance to (ii) rectify design and architecture issues is an effective way for doing this kind of work in practice. It helps to jump up a level from code smells to design problems and their rectification.       

\section{Case Overview}

The case company under study is a software startup whose product is a customer satisfaction surveying and response analysis tool for companies. The main product is a {\em dashboard} where survey responses and data analysis are visualized for the customers.

The application has been under active development for five years, with the team growing from one hired consultant, who did the initial implementation, to six in-house developers. During this time, the company has followed the principles of lean startup \cite{ries2011lean} ensuring the delivery of features and a product that has a good fit in the market. Over the years, the company has made numerous pivots, or changes in product strategy but not in product vision, particularly during the first year of the development.

The software product was composed of the main backend and frontend (customer dashboard). In addition, there were six other services that dealt with sending the surveys, giving the responses to the main backend, and performing some analysis on the responses. All of the services were hosted in Amazon Web Services (AWS) \cite{aws}.

Here, the focus is placed on the main backend, built with Clojure during the first year of development. It was a simple web application with a database. It not only contained the API for the frontend to use the data in its database, but some background tasks like an importer that brought in the responses of the surveys from a third-party tool used to send the surveys. Although being in the same Leiningen \cite{leiningen} project, the importer was run in its own AWS Elastic Beanstalk instance through an environment variable. In addition to these two independent applications, the same project contained an admin dashboard done with ClojureScript \cite{clojurescript}. This admin dashboard and the importer code had not been changed after the second year of development.

Most of the code written from the second year of development onward was written in Java. To achieve this, the only developer at that point built an interoperability helper to be used as a bridge between Clojure and Java code. This helper file had not been changed since the second year of development. This helper helped with the translation of objects between Clojure and Java. What this meant in practice is that every single object passed to the Clojure side was of the Java Map type. The MapHelper class was created in Java to make it easier to make the transformations of objects into Maps. As a consequence of the need to interoperate with Clojure, another special characteristic of the Java code was the use of classes with only static methods. This was due to Clojure being a functional programming language and thus not supporting classes and objects. This made the Java code very untypical and in consequence difficult to write and read with an idiomatic Java mindset. In total the Java code extended through about 20.000 lines of code, making it a small software system.

Finally, testing was clearly a problem. The Clojure part had 23 tests in 5 files, all written during the first months of the development. These were run on every build, but their coverage was low. Furthermore, nothing was really modified in the Clojure code -- adding and modifying endpoints of the web application, which was the main development target, did not require this -- and these tests would always pass. On Java side, there were 505 tests in 55 files, but 120 tests were marked as ignored and thus would not be run, unless the annotation used for ignoring them was removed. Java tests were not automated; they  were never a part of the official verification of new releases, as they could only be run manually in the IDE.

\section{From Code Smells to AntiPatterns}

To find out all the things that needed changing concerning the product, an analysis of both the codebase and the development practices was performed. The intention was to find the underlying AntiPatterns that could be hiding behind very apparent code smells that had already surfaced in the development. Then, the AntiPatterns would provide a systematic way to fix the identified problems.

\subsection{Sniffing for Code Smells}

When working on a codebase that developers are very familiar with, developers might lose their sense of smell, hence making them unable to notice the most obvious code smells. Furthermore, haste and time pressure can worsen the situation, as the attention of developers is focused on the tasks at hand and not the codebase as a whole. This was also the case with this product; the need to build new features and fix critical bugs impeded developers from taking a wider view to understand the underlying problems.

To avoid possible biases resulting from familiarity with the code, the decision to use static code analysis tools was made. Two static code analysis tools were used in this case study, CodeMR \cite{codemr} and IntelliJ IDEA's code inspection tool \cite{codeinspection}. These tools have complimentary features -- CodeMR has a more visualization centric approach that measures the quality of code with different metrics on the Java class level, while IntelliJ IDEA's code inspection tool shows concrete potential coding problems line by line. Unfortunately, the tools did not support both Clojure and Java, and hence only the Java code was analyzed. Furthermore, a decision was made that the Clojure code should be removed in the codebase in short-to-medium term in any case. Both tools were used in their default configuration without tuning the threshold values nor the problem categories used in reporting problems. 

\subsubsection{CodeMR}

CodeMR is a software quality tool that supports multiple languages, with integrations to multiple IDEs. It provides an insight to the quality of software through an array of attributes -- coupling, complexity, cohesion, size, and so on. In general, these metrics are often affected by several features in code, which makes them promising for seeking code smells.

The Java code of the backend of the application under study was analyzed with the help of the CodeMR IntelliJ IDEA plugin version 2020.4.1-release-2020.2. Test files were ignored for this analysis. The analysis showed that the number of code lines is nearing 20.000, indicating that the size itself might help hiding details and making it more complex to find something specific a developer might be looking for. The number of classes and packages was reasonable or even low, with 21 packages and just under 19 classes per package on average, totalling 397 classes. The number of external packages and classes is a bit more concerning -- 92 and 374, respectively -- as the number of external classes is nearly the same as the native ones. While using external libraries is a standard procedure in today's software development, a review of the external dependencies would be advisable based on these numbers. Finally, CodeMR had identified that 34 classes were problematic and 1 was highly problematic.

\begin{figure}[t]
    \centering
    \includegraphics[width=\columnwidth]{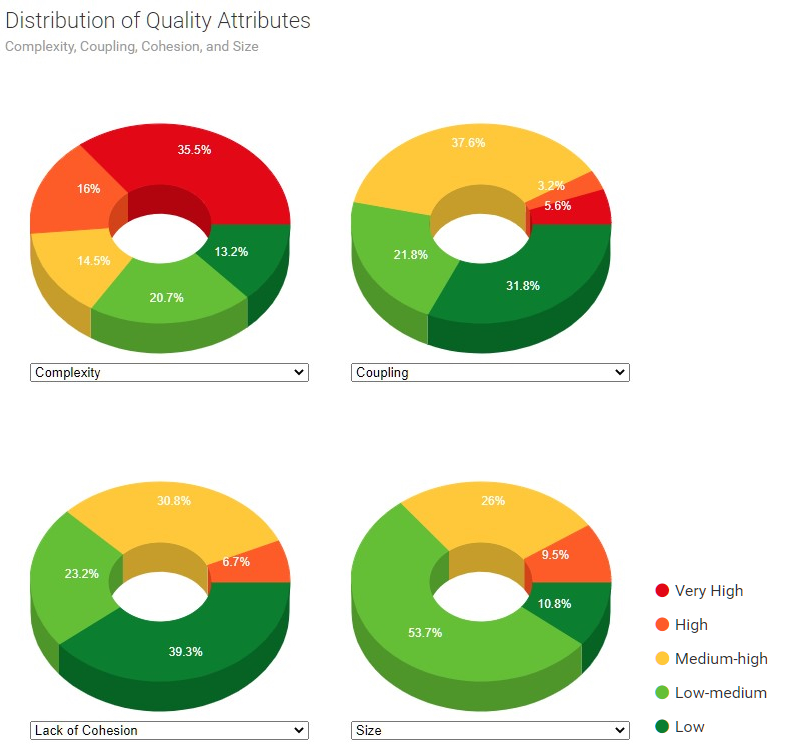}
    \caption{CodeMR metrics from the backend analysis.}
    \label{fig:codemrmetrics}
\end{figure}

As for the four key metrics discussed above -- coupling, complexity, cohesion and size -- CodeMR provided a chart view (Figure \ref{fig:codemrmetrics}) where each chart visualizes the amount of classes with high or low values of the metric. Based on the charts, complexity and coupling seem to be the biggest possible problems. According to CodeMR, 35.5\% of the code has the highest level of complexity, and only just over a third of the code was categorized as having medium to low complexity. This was confirmed by the feeling of developers that code is difficult to reason about in many places. For coupling, just under half of the code was deemed to have medium to very high coupling. This was also confirmed by developers who saw that modifications often cascaded to a flood of changes all over the codebase. Lack of cohesion and size seemed to do a bit better. In both respects, well over half of the codebase was identified to have medium to low levels of both metrics, and no instances of very high levels were found. The experiences of the developers again match the result.

\begin{figure*}[t]
    \centering
    \includegraphics[width=1.85\columnwidth]{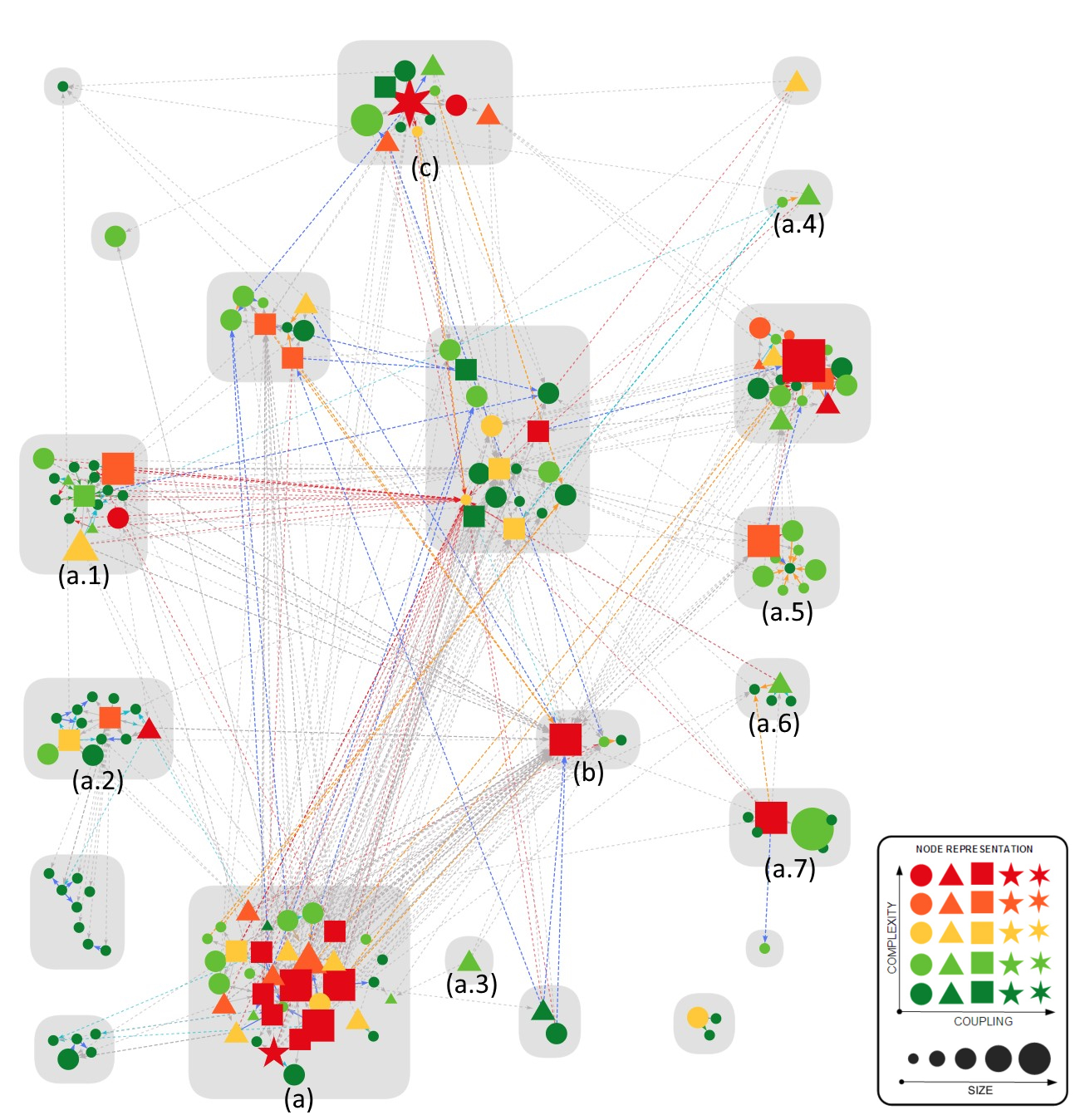}
    \caption{CodeMR graphic showcasing the relationship between classes in their respective Java packages.}
    \label{fig:codemrpackages}
\end{figure*}

For a more detailed inspection of the codebase, CodeMR also provided a variety of graphs and detailed metrics for each class. Figure \ref{fig:codemrpackages} shows all the classes analyzed in their respective packages, using colors to indicate complexity, shapes for coupling, the size of the symbols indicating size, and  cohesion with a web of dependencies between the symbols. This more detailed view helps to understand and pinpoint the classes that would be potential problem makers. Some of the packages have been marked with letters for further reference below.

The first indication in the graph is that there were some classes that seemed to be used all over the codebase. One example is a class in package (b), which was used to do every single database query, so it was used by all service classes of the codebase. Most classes in package (a) and its subpackages (a.1-7) had lots of dependencies, which might be reasonable. These packages contained all the main service classes that mostly deal with the business logic of the software. Another thing that was apparent from the graph is that although the average size of classes was just over 43 lines of code, the distribution of real code lines was very uneven. There was a considerable number of classes that clearly were either too big or were nearing that limit. Based on the detailed data provided by CodeMR, there were 11 classes with more than 300 lines of code, with the biggest class that can be seen as a big red star on package (c) having 950 lines of code. With the exception of package (a) that had many more complex and coupled classes, most packages seemed to have one particularly complex and coupled class. This seemed to suggest that there was a bloated main class and too few helper classes around it. 

To summarize, CodeMR's analysis provided tangible evidence that there were areas of the codebase that might have or cause problems. In general, this was what the developers were expecting, to a large degree.

\subsubsection{IntelliJ IDEA code inspection}

IntelliJ IDEA offers static analysis of the code in real time, while the developer writes it. This feature, called code inspection, can be run for a given part of the project or for the full codebase. In comparison to the static code analysis made with CodeMR, IntelliJ IDEA's code inspections try to give warnings regarding concrete problems in the code, line by line. The tool even suggests fixes that can be automatically applied, thus automating refactoring for common problems. Running the inspection tools for the whole Java codebase of the backend under study on version 2020.2.2 of IntelliJ IDEA resulted in a detailed report regarding problems in code.

The 533 typos found by the tool under the proofreading category were for the most part caused by the use of acronyms and words that the dictionary simply did not recognize. However, some of the warnings revealed real typos in the names of variables and method names that just had not been noticed. 

The Java category contained most of the warnings the tool reported of which the most meaningful ones have been collected into Table \ref{table:ideawarnings}. The first thing that was noticed was the number of declaration redundancy warnings. Over half of all the warnings were of this kind. Looking closer into the subcategories, 1489 of those warnings were unused declaration warnings. This was attributed to using Clojure as the entry point to the application, which the tool was unable to deduce.

\begin{table}[t]
    \centering
    \caption{Most meaningful warnings under the Java top level category given by the IntelliJ IDEA Code Inspection.}
    \label{table:ideawarnings}
    \begin{tabular}{p{0.18\columnwidth}|p{0.45\columnwidth}|p{0.14\columnwidth}}
    Category
    & Subcategory
    & Instances 
    \\
    \hline \hline
    Control flow issues
    & Pointless boolean expressions
    & 57
    \\ \hline
    Data flow
    & Redundant local variable
    & 75
    \\ \hline
    Declaration redundancy
    & Declaration can have final modifier
    & 140
    \\ \hline
    Declaration redundancy
    & Unused declaration
    & 1489
    \\ \hline
    Performance
    & String concatenation in loop
    & 42
    \\ \hline
    Probable bugs
    & Constant conditions \& exceptions
    & 99
    \\ \hline
    Probable bugs
    & Unused assignment
    & 76
    \end{tabular}
\end{table}

In addition, there were lots of different kinds of warnings. Those that seemed more like repeated code smells rather than isolated mistakes included the use of pointless boolean expressions, redundant local variables, string concatenations in loops, constant conditions \& expressions, and unused assignments. Except for string concatenations in loops that might negatively affect performance, the rest mostly affect the readability and the clarity of intention of the code. In the case of the constant conditions and expressions, most of them were clearly potential bugs that could end up triggering null pointer exceptions.

\subsection{Confirmed Code Smells}

The static code analysis tools gave some good insight on the possible location of the most obvious problems in the case project. The likely code smells that could be inferred by studying the results produced by both tools were: Mysterious Name, Mutable Data, Divergent Change, Shotgun Surgery, Feature Envy, Temporary Field, Large Class and Data Class. These findings then required a manual inspection to corroborate them and to see how pervasive they might be. The inspection could also uncover other problems and code smells that the static code analysis tools might have overlooked.

For example, the web of dependencies shown in Figure~\ref{fig:codemrpackages} (CodeMR) suggested to the inspector the whiff of Divergent Change, Shotgun Surgery and Feature Envy. The inspector then studied the suspect classes in more detail (illustrative examples of analysis are given below). Some smells were more directly observable in the tools' output, like Large Class (CodeMR) and Temporary Field (IntelliJ).

Manual analysis was also done to create a priority list of the most complex classes, in the order of severity and amount of warnings. Next, we discuss three representative classes in more depth, explaining the findings in them, and finally considering the codebase as a whole.

\subsubsection{Code smells in Class 1}

Class 1 was considered as the most problematic by both static code analysis tools. The class processes a configuration and sends a screenshot of the application's dashboard as email to the users of this particular configuration. This task has four parts: Process the configuration, retrieve the dashboard, create an image of the dashboard, and send the email with the image as an attachment. 

As the class was single-handed responsible for doing all this, with no helper classes, its size was considerable. Even the main function, used to initiate the process, was 91 lines of code, with a very complex conditional flow that was nested five levels deep. Several other functions were also well over 50 LoC. Furthermore, there was another class inside the main class instead of being in its own file. This internal class was potentially even more problematic; some of its functionality was not implemented as its own methods but by those of the parent class. For example, setters for some fields of the class were outside the class itself.

A further suspicious factor was passing the same parameters from one method to another, which was  related to using static methods. For instance, a data source object used to connect to the database was passed from the very first entry point to wherever database connections were needed. Consequently,  parameter lists were very long, with some inconsistency in the location of the different parameters. This was harmful for readability of the code.

In summary, Class 1 is definitively smelly, containing at least the Long Function, Long Parameter List, Mutable Data, Divergent Change, Data Clumps and Large Class smells.

\subsubsection{Code smells in Class 2}

Class 2 is used to define filtering of the data requested in a query. It contains a set of parameters that can be transformed directly into SQL query parameters. It performs three tasks, parse itself from JSON, transform into an SQL query, and serialize itself into JSON.

An obvious smell in the class is the extensive use of primitive types to represent all the data in this one class instead of defining subclasses encapsulating more cohesive groupings of the data, which would simplify the design. The rationale for this solution is the use of a custom serializer and deserializer for JSON. The serializer was built using Java StringBuilder, whereas, the deserializer was implemented with JSON in Java's JSONObject interface. To add to the complexity, the class had also annotations used to serialize and deserialize objects into JSON through the FasterXML Jackson library \cite{jackson}, which seemed redundant due to the extensive use of the custom implementation throughout the codebase.

In summary, Class 2 has somewhat comprehensive smells. These include Long Function, Divergent Change, Primitive Obsession and Large Class.

\subsubsection{Code smells in Class 3}

Class 3 handles all the different possible configurations used for the dashboards. There are dozens of different configurations, and more are added whenever new features require them. The main responsibility of the class is to create, read, update and delete the different kinds of configurations in the database.

The class starts with a list of 18 different SQL string constants, all written in different styles. Most are one-liners, with varying use of upper and lower case letters, but two of them are divided into several lines to improve readability. Some of these SQL statements use capitalized SQL keywords, others do not, and some mix the styles in the same statement. The class contains 10 additional internal classes, resulting in a complex design. Except for one of the many methods in the class, they have a reasonable size. However,  many methods use temporary variables just before the return statement, which seems confusing.

The smells in this class include Long Function, Divergent Change and Large Class.

\subsubsection{Summary}

The three examples above showed some of the particularities of the case project, many of which are tangled in all classes. Furthermore, change-related smells, Divergent Change and Shotgun Surgery, were well represented. There were numerous instances of Divergent Change on larger classes, as they had been modified for a number of reasons. Furthermore, passing parameters  through all the method calls had made the codebase prone to Shotgun Surgery because if even one new parameter was needed in a method deep down a call chain, the whole chain would need to be modified, which implies changes in many classes. Another recurring problem was the use of public fields, which is an instance of both the Global Data and Mutable Data smells. The origins of this problem was tracked to the design of the interaction between Java and Clojure, where all fields that Clojure accessed had to be public.

There was also a lot of utility functionality embedded in service classes, which were then used by other service classes, presenting a clear case of Feature Envy. Sometimes this created circular dependencies that were only hidden by the fact that the code is mostly composed of static methods. This would become a real problem once the classes would be transformed to follow object-oriented idioms more closely.

To summarize, the code smells that the static code analysis tools had hinted to were confirmed in manual inspection. In addition to those code smells, the manual analysis also revealed some smells that could not be inferred from the reports of the automated tools. Hence, various tools can help find some problems, but a manual review is able to find more subjective smells.

\subsection{Identifying AntiPatterns\label{underlyingantipatterns}} 

Although many problems in the codebase were identified through the search for code smells, the codebase seemed to have deeper problems than the surface level smells alone suggested. Most code smells are usually centered on one small section of the code and can appear repeatedly, but they are usually easily fixable with the use of simple code refactorings. In search of deeper recurring problems, the codebase was further inspected manually using the smelly sections of the codebase as the starting point to study if any AntiPatterns \cite{brown1998refactoring} were present. For conciseness, we use here a truncated form of the full AntiPattern template with the addition of the 'Related Code Smells' field to signify the code smells that led to the discovery of the presence of the AntiPattern in the codebase:

\begin{description}
    \setlength{\itemsep}{0pt}
    \setlength{\parskip}{0pt}
    \item[AntiPattern Name:] AntiPattern's name, variant names
    \item[Refactored Solution Name:] Identifying name of the refactored solution
    \item[Root Causes:] List of the root causes pertinent to the AntiPattern
    \item[Related Code Smells:] Code smells that might be symptomatic of the AntiPattern which were identified through the static code analysis
    \item[General Description:] An overview of the AntiPattern, its forces and characteristics.
\end{description}

\noindent
The identified AntiPatterns are listed below. Each AntiPattern is followed by some concrete examples of how they were present in the system. The initial plans and actions to be taken to solve them are also presented were applicable. In addition, a link to the related code smells is also given.

\subsection*{Lava Flow AntiPattern}

\begin{description}
    \setlength{\itemsep}{0pt}
    \setlength{\parskip}{0pt}
    \item[AntiPattern Name:] Lava Flow, Dead Code \cite{brown1998refactoring}
    \item[Refactored Solution Name:] Architectural Configuration Management
    \item[Root Causes:] Avarice, Greed, Sloth
    \item[Related Code Smells:] Temporary Field
    \item[General Description:] This AntiPattern is characteristic of software that has gone through lots of changes in direction. Each change in direction generates new code that is then never removed when the idea behind it is abandoned. This results in sections of code that might not clearly be no longer used and that never are removed. This adds to the complexity of the code. Parts of the code might even end up being reused elsewhere, making the dependency structure and the elimination of this dead code even harder.
\end{description}

This AntiPattern is elusive, as developers need to look for it with intention, and problems are rarely found by coincidence. Things like unused variable and temporary fields might be an indication of its existence, but in most cases the code will look just normal, but it just implements functionality that is no longer used. As already mentioned, the case project underwent many changes in direction early on in the development. In this process, code that was no longer used was left in the codebase, making it obvious that this AntiPattern was met. 

A particular detail of this AntiPattern is related to characteristics of web software. In general, knowing if the endpoints of a web application are actively used cannnot be verified from its code alone, but an external analysis must be done. This was carried out in the case project. The code of all the client services that used its endpoints were searched for the usage of the endpoints. In addition to this, AWS CloudWatch Logs Insights \cite{cloudwatchinsights} tool was used to count the number of http requests done to the endpoints from the http logs of the production servers. The result of this analysis was that 77 endpoints from over 350 were no longer used at all. These were subsequently removed, resulting in the deletion of 2690 lines of dead code.

Another symptom that might indicate of the presence of this AntiPattern is the existence of columns and tables in the database that are either completely empty or always have the exact same value. Some instances of this were found in case project. They were immediately removed if no refactoring was needed; if refactoring was considered necessary, associated cleaning tasks were added to the project backlog.

\subsection*{Functional Decomposition AntiPattern}

\begin{description}
    \setlength{\itemsep}{0pt}
    \setlength{\parskip}{0pt}
    \item[AntiPattern Name:] Functional Decomposition, No Object-Oriented Pattern \cite{brown1998refactoring}
    \item[Refactored Solution Name:] Object-Oriented Reengineering
    \item[Root Causes:] Avarice, Greed, Sloth
    \item[Related Code Smells:] Long Parameter List, Global Data, Mutable Data, Data Clumps, Data Class
    \item[General Description:] This AntiPattern is in its simplest term the misuse of an object-oriented language as it were a functional or structural language. This can make the code very convoluted. object-oriented programming has lots of beneficial design patterns which this kind of code does not take advantage of. Depending on how widespread this AntiPattern is the reengineering work needed to fix it might be overwhelming and must be done in increments.
\end{description}

\noindent
Due to the Clojure roots of the project, parts written in Java did not follow the usual object-oriented structure of classes that are instantiated into objects. Instead, the code relied on the use of static methods that could be run without creating new instances. In addition to the extensive use of static methods, smells like long parameter lists, global data and Mutable Data were among clear indicators of the presence of this AntiPattern.

As this was a very overarching AntiPattern in the codebase, it would be no easy task to convert the whole codebase into beautiful object-oriented code. As it had already been decided that the original Clojure code would be eliminated, the developers decided to refactor the system to follow the inversion of control patterns of Java Spring \cite{johnson2009professional} when migrating it to a new Spring MVC based repository. When migrating a part of the code, it was also noticed that the current paradigm had made circular dependencies in the code possible. These would start to surface when instantiating the classes instead of using their methods statically. Another detail associated with this refactoring is passing parameters through function calls -- services containing the utility and necessary information could be directly included in objects that need them. Work refactoring this AntiPattern out will continue in tandem with the removal of the old Clojure code which had been started.

\subsection*{Spaghetti Code AntiPattern}

\begin{description}
    \setlength{\itemsep}{0pt}
    \setlength{\parskip}{0pt}
    \item[AntiPattern Name:] Spaghetti Code \cite{brown1998refactoring}
    \item[Refactored Solution Name:] Software Refactoring, Code Cleanup
    \item[Root Causes:] Ignorance, Sloth
    \item[Related Code Smells:] Long Function, Divergent Change, Shotgun Surgery, Loops, Message Chains, Insider Trading, Large Class
    \item[General Description:] Spaghetti code needs no introduction, being the most famous of all AntiPatterns and known even to developers that are not aware of the concept of AntiPatterns itself. This AntiPattern depicts code without a logical structure. This kind of code will be difficult to understand even to the original developer if it has not been in its focus even for a few weeks. The code might contain large classes, with convoluted flows in a single function. Some classes might also contain clearly out of place functionality that then other classes might use, making the dependencies between classes very difficult to understand.
\end{description}

\noindent
In the case project, large classes and long functions found with static code analysis were clear examples of spaghetti code. These classes had one main entry point function that would then go through a convoluted flow of conditional expressions and function jumps that made the code difficult to understand. Although these classes were clearly in need of refactoring, the developers rarely had the time to refactor them while making changes in them. 

Another instance of this AntiPattern was the misplacement of functionality in classes where the functionality clearly did not belong to. The rationale for this was keeping the number of classes at a minimal level, by including all the functionality needed by a class in itself. This was done for the sake of development speed, knowing that some functions would better fit another service or a helper class, which could then be used by other classes, too. Ultimately, the solution would have been acceptable, assuming refactoring would take place once a function functionality embedded into the class was needed by other classes. Unfortunately, this was not the case, resulting in bloated classes and weird dependency graphs between them.

To fully eliminate spaghetti code, a change in the mentality regarding when to refactor was needed, making refactoring a first-class citizen in the development workflow. Then, instead of forcing new changes to code that clearly is in need of refactoring, the code is refactored first to make the change easier to implement and understand. This piecemeal approach can be complemented with bigger refactoring sessions, when time was available. A few instances of this AntiPattern were soon opportunistically refactored at the same time as the classes needed other functional changes. The remaining cases were added to the backlog for later refactoring.

\subsection*{Reinvent the Wheel AntiPattern}

\begin{description}
    \setlength{\itemsep}{0pt}
    \setlength{\parskip}{0pt}
    \item[AntiPattern Name:] Reinvent the Wheel, Design in a Vacuum, Greenfield System \cite{brown1998refactoring}
    \item[Refactored Solution Name:] Architecture Mining
    \item[Root Causes:] Pride, Ignorance
    \item[Related Code Smells:] Duplicated Code
    \item[General Description:] This AntiPattern refers to the lack of reuse of code from the codebase itself or of external available libraries. It usually stems from the lack of knowledge of what has previously been done or the believe of some developers that think that they could do it better themselves.
\end{description}

\noindent
This is the only AntiPattern we encountered that had no code smells to even slightly indicate its presence -- there was no true duplicate code.  Therefore, to spot this AntiPattern a logical view of what was readily available either in the codebase itself or in open external libraries.

The first example of wheel reinvention in the case project was multiple implementations of comma-separated values (CSV) file parsing and writing. Although an external library had been introduced for this, only some of the instances of handling CSV files used this external library. Others created their own, non-trivial specific implementations that were not reused nor reusable by other classes. This was so widespread yet low risk issue, that fixing it by harmonizing the code was added to the backlog as a low priority task.

A similar instance of wheel reinvention is related to JSON serialization/deserialization. This was sometimes done by hand, instead of relying on libraries that are readily available. JSON is a much more complex format than CSV, hence making its manual implementation more risky. At the same time, an external library was used in the case project. As for refactoring, it was decided that whenever manual JSON handlers need updates, the external routine will be used instead.

\section{Discussion}

Working on the case project has provided us confidence that the combination of code smells and AntiPatterns are a practical way to manage technical debt. 

In the case project, (i) areas of code with code smells did often have underlying AntiPatterns. Each AntiPattern usually had its own set of code smells that were related to it but there was no direct causation the other way around. Instead, code smells served as indicators of the need for deeper inspection of the underlying behavior. Furthermore, (ii) the use of AntiPatterns to analyze and fix the problems in the codebase was very helpful. Looking for AntiPatterns instead of individual problems made it easy to find more overarching patterns and thus using the refactored solution of the AntiPattern enabled planning and making fixes that would improve the codebase’s structure and readability as a whole. 

Another observation and confirmation made was the added difficulty that not having tests causes for dealing with technical debt in a codebase. Refactoring, the main tool used to improve code without changing its behavior, relies heavily on having tests that can confirm that regressions are not introduced while trying to fix other problems. Adding tests is thus always the first step towards getting rid of the accrued technical debt. Furthermore, some AntiPatterns were continuously reintroduced even after having realized their existence and its active removal in other areas of the code. It is thus of the utmost importance to understand AntiPatterns, particularly those that have already been previously identified, and have good peer review practices that make it possible to catch them before they end up in the production code and before not too long the project is back on square one.

Numerous directions for future work in the field of this paper exist. Firstly, since the goal of this paper is to summarize our first-hand experiences, a direct successor of this work would be a more comprehensive study of advanced static code analysis tools that could aid in the identification of AntiPatterns in industrial projects. Another direction of further research is the matter of scaling the approach in terms of the size of the codebase. To confirm central findings of this paper, similar studies in the context of different kinds of software development projects and types of companies are needed to hypothesize if the use of AntiPatterns in technical debt management is feasible on projects of different sizes and types. Finally, a more constructive research theme would be to build a repository of AntiPatterns from previous, related projects with concrete examples of refactored and implemented solutions. This in turn would help understanding possible problems in the new context and the extent of remedies required.

\section{Conclusions}

To conclude, understanding AntiPatterns has the potential to help projects alleviate the technical debt that they have accrued. However, the concept of AntiPatterns is not as widely known or used with their full meaning, including their refactorings, as that of design patterns, and hence it is less common to analyze software systems with them. 

In this paper, we presented our experiences in detecting and removing technical debt from a case system where the codebase had clear signs of decay. In our experience, the concepts of code smells and AntiPatterns considerably helped in the process. Furthermore, while tools for performing static analysis did help in the process, also human insight and manual work were deemed important, in particular when performing a deeper analysis of implications of the problems.


\bibliographystyle{IEEEtran}
\bibliography{bib}

\end{document}